\documentclass[useAMS,usenatbib,usegraphicx]{mn2e}
\usepackage{amsmath,color}
\usepackage[latin1]{inputenc}
\usepackage{JournalNames}
\newcommand{\vel}{\textnormal{v}}
\newcommand{\half}{\ensuremath{
    \frac{1}{2}}}

\def\tens#1{\ensuremath{\mathsf{#1}}}
\begin{document}

\title{Local Turbulence Simulations for the Multiphase ISM}

\author[R. Kissmann et al.]{R. Kissmann,$^1$
J. Kleimann,$^{2,3}$
H. Fichtner,$^4$
and R. Grauer,$^5$\\
$^1$ Institute for Astronomy \& Astrophysics, University of T\"ubingen, Auf der
 Morgenstelle 10, 72076 T\"ubingen, Germany\\
$^2$ Max-Planck-Institut f\"ur Sonnensystemforschung, 37191 Katlenburg-Lindau, 
Germany\\
$^3$ Auroral Observatory, University of Troms\o, 9037 Troms\o, Norway\\
$^4$ Institute for Theoretical Physics IV, Ruhr-University Bochum,
44780 Bochum, Germany\\
$^5$ Institute for Theoretical Physics I, Ruhr-University Bochum,
44780 Bochum, Germany}

\date{Accepted 2008 September 17. Received 2008 September 16;
  in original form 2008 March 18}

\pagerange{\pageref{firstpage}--\pageref{lastpage}} \pubyear{2008}

\maketitle

\label{firstpage}

\begin{abstract}
  In this paper we show results of numerical simulations for the
  turbulence in the interstellar medium. These results were obtained
  using a Riemann solver-free numerical scheme for high-Mach number
  hyperbolic equations. Here we especially concentrate on the physical
  properties of the ISM. That is, we do not present turbulence
  simulations trimmed to be applicable to the interstellar medium. The
  simulations are rather based on physical estimates for the relevant
  parameters of the interstellar gas.
  Applying our code to simulate the turbulent plasma motion within a
  typical interstellar molecular cloud, we investigate the influence
  of different equations of state (isothermal and adiabatic) on the
  statistical properties of the resulting turbulent structures. We
  find slightly different density power spectra and dispersion maps,
  while both cases yield qualitatively similar dissipative structures,
  and exhibit a departure from the classical Kolmogorov case towards a
  scaling described by the She-Leveque model.
  Solving the full energy equation with realistic heating/cooling
  terms appropriate for the diffuse interstellar gas, we are able to
  reproduce a realistic two-phase distribution of cold and warm
  plasma. When extracting maps of polarised intensity from our
  simulation data, we find encouraging similarity to actual
  observations. Finally, we compare the actual magnetic field strength
  of our simulations to its value inferred from the rotation
  measure. We find these to be systematically different by a factor of
  about 1.5, thus highlighting the often underestimated influence of
  varying line-of-sight particle densities on the magnetic field
  strength derived from observed rotation measures.

\end{abstract}

\section{Motivation}
Interstellar space is filled by a dilute partially ionised gas called
the Interstellar Medium (ISM). In contrast to the classical belief
that it is a more or less undisturbed multi-phase medium existing in
pressure equilibrium, the ISM is nowadays thought to be in a turbulent
state \citep[see, e. g.,][]{ElmegreenScalo2004}. There are several
hints for this turbulence coming from observations as are for example
the broadening of spectral lines, the chemical mixing of the ISM and,
indirectly, also the star formation rate \citep[see,
  e. g.,][]{ElmegreenScalo2004,ScaloElmegreen2004}. While turbulence
in the latter case was accounted for simply by considering a turbulent
pressure, a global description of the ISM has to take the turbulence
directly into account \citep[see,
  e. g.,][]{Larson1981,LizanoShu1989,Elmegreen1993}.

While turbulence is certainly important on the largest scales, also
atomic particles are thought to be strongly influenced by the
turbulent fluctuations: The variability of the gas and of the
electromagnetic field within are the means by which cosmic ray
particles are scattered during their motion through the Galaxy and the
intergalactic medium. The properties of the turbulence needed to
calculate the diffusion lengths and similar quantities for these
particles can, however, not be directly obtained from observations.
Lacking the possibility to bring a probe into interstellar space in
the near future \citep[see, e. g.,][for the far-future
prospects]{FichtnerEtAl2006}, observations using different parts of
the electromagnetic spectrum are the only method to gain such
information. So far, however, the observational access to fundamental
turbulent quantities in the ISM is very limited.
Due to the fact that also analytical models for turbulence are still
lacking -- at least for a system as complex as the ISM -- we have to
rely on numerical means to gain some insight into interstellar
turbulence. Fortunately, the advance of computational resources allows
even such complex systems as the ISM to be modelled numerically in the
fluid description. It has, however, to be kept in mind that the ISM is
not just an arbitrary plasma laboratory for the turbulence
scientist. Even though the ISM shows a wide range of parameters, these
parameters are interrelated in characteristic ways. Therefore, one has
to properly prescribe the physical parameters of the corresponding
simulations. In contrast to basic turbulence research simulations for
example, one has to choose the external driving in accordance with the
observations. Also heating and cooling processes usually have to be
taken into account. In this connection numerical simulations are an
interesting method to derive the influence of turbulence on the
temperature of the ISM.

An additional complication arises by the fact that most parts of the
ISM are highly compressible, thus allowing for the possibility of
shocks to form in the turbulent plasma \citep[see,
  e. g.,][]{BoldyrevEtal2002}. With this and the requirement for high
spatial resolution, the demands on a numerical solver are very
high. For this work we used an implementation of a conservative,
central finite-volume scheme, which we demonstrate to be very suited
for the complex task at hand. By way of this scheme we perform
simulations for the ISM turbulence, which are adapted to the
interstellar medium.

The structure of this paper is as follows. First, we discuss the
physical description of the turbulent ISM. In the subsequent section
we introduce the scheme used for this modelling.  Finally, we show
first results of the code's application to interstellar turbulence.

\section{The Physical Model}
\label{equations}
In this work we concern ourselves with the structure and fluctuations
of the ISM. For this we first need to find a suitable physical
description. Naturally, we will not consider a model describing the
individual particles when dealing with the large gas masses in
interstellar space. Even though the gas is very dilute, we will
consider such large volumes that such a description is not
necessary. Due to the limited computational resources, we will also
not use a kinetic description. Therefore, we will rely on a fluid
description for the interstellar plasma.

Consequently the ISM will be described by the ideal MHD equations:
\begin{eqnarray}
  \frac{\partial \rho}{\partial t} + \nabla \cdot \left(\rho \vec{u}\right)
  &= 0
  \\
  \frac{\partial \left(\rho \vec{u} \right)}{\partial t}
  + \nabla \cdot \left(\rho \vec{u} \vec{u} \right)
  + \nabla \cdot \tens{P}
  + \frac{1}{\mu_0}\vec{B} \times \left(\nabla \times \vec{B}\right)
  &=
  0
  \\
  \label{EqInduction}
  \frac{\partial \vec{B}}{\partial t}
  +\nabla \cdot \left(\vec{u}\vec{B} - \vec{B}\vec{u}\right)
  &=
  0
  \\
  \label{EqEnergy}
  \frac{\partial e}{\partial t}
  + \nabla \cdot \left(
  \left(e + \frac{B^2}{2 \mu_0} + \tens{P}\cdot \right)\vec{u}
  - \frac{1}{\mu_0}\left(\vec{u}\cdot\vec{B}\right)\vec{B}
  \right)
  &= \\
  \nonumber\qquad\qquad\qquad\qquad\qquad
  n_{\mathrm{H}}  \Gamma - n_{\mathrm{H}}^2 \Lambda
  + \nabla \cdot \left(\kappa \nabla T \right)
\end{eqnarray}
The relevant quantities are the mass density $\rho$, the flow velocity
$\vec{u}$, the magnetic induction $\vec{B}$, and the overall energy
density $e$. For the pressure tensor $\tens{P}$ we assume a
description via a scalar pressure to be sufficient, thus, yielding
$\tens{P} = p \hat{1}$, where $\hat{1}$ is the unit matrix. The source
terms of the energy equation are the heating ($\Gamma$) and cooling
functions ($\Lambda$), which both depend on the density and the
temperature -- here $n_{\mathrm{H}}$ is the hydrogen number
density. Finally $\kappa$ is the constant coefficient of heat
conduction, which was only included, when also heating and cooling
terms were present to avoid unresolved growth of the thermal
instability on the grid scale
\citep[see][]{KoyamaInutsuka2004ApJl}. This was chosen as $\kappa =
5\cdot 10^{-4}$ in normalised units.

The above equations are the \emph{compressible} form of the MHD
equations -- allowing for the formation of shock structures as they
are observed in the ISM \citep[see, e. g.,][]{vestuto_etal_2003}. As
an additional condition, we also have to fulfil
\begin{equation}
  \nabla \cdot \vec{B} = 0
\end{equation}
at all times of the simulation. With these equations we possess the
general framework for the simulation of the interstellar plasma. As
mentioned above, however, there are several issues which have to be
taken into account in addition to the fluid description of the plasma.

\subsection{Implications of the Phase Structure}
One of the most important issues setting the ISM apart from a
turbulence laboratory is connected to the classical concept of its
phase structure. From observations, it is known that the ISM consists
of at least three distinct phases, being brought about by unstable
regions in the pressure curve for the ISM. These result from an
equilibrium of heating and cooling processes in the ISM. The resulting
stable phases are characterised by some typical density and
temperature (which, however, might vary quite strongly). This
structure also allows for two basically different simulation
scenarios.

On the one hand one could try to capture the multi-phase interstellar
medium, whereas on the other hand only a small volume of the ISM could
be taken into account, which would then reside exclusively in one of
the phases. The latter option, thus, allows for a simplification of
the above system of equations. The simplest possibility to describe
one of the phases would be to investigate an isothermal medium, which
just possesses the average temperature of the corresponding phase. In
this case the evolution equation for the energy density
(\ref{EqEnergy}) can be replaced by the following equation of state in
order to close the system of equations:
\begin{equation}
  p = c_{\mathrm{iso}}^2 \rho
  \qquad \textnormal{with} \qquad
  c_{\mathrm{iso}} = \sqrt{\frac{k_{\mathrm{B}} T_0}{m_{0}}}.
\end{equation}
Here $k_{\mathrm{B}}$ indicates the Boltzmann constant, $m_{0}$ is the
reduced mass of a particle of the ISM, and $T_0$ is the temperature of
the corresponding phase.

When investigating several phases at the same time, use of the full
energy equation is mandatory. To allow for the existence of the phase
structure, in this case, we have to include the corresponding heating
and cooling functions for the ISM. For temperatures up to $10^{4.2}$ K
we use the most general form for the cooling function given as:
\begin{equation}
  \label{EqHeating}
  \Lambda(x,T) = \sum_i \frac{n_{X_i}}{n_H}
  \left[x \lambda_{\mathrm{e}}(X_i,T) + \lambda_{\mathrm{H}}(X_i,T) \right]
\end{equation}
Here the individual cooling functions $\lambda_{\mathrm{H}}(X_i,T)$
and $\lambda_{\mathrm{e}}(X_i,T)$ are given for the individual species
$X_i$ in \citet{Penston1970} and \citet{DalgarnoMcCray1972}
analytically. The resulting cooling rate depends on the fractional
ionisation $x = n_{\mathrm{e}}/n_{\mathrm{H}}$ and the relative
elemental abundances $n_{i}/{n_{\mathrm{H}}}$, which were also taken
from \citet{DalgarnoMcCray1972}. For higher temperatures we use the
approximate cooling function suggested in \citet{GerritsenIcke1997}
For the element abundances we use the \emph{depleted values} from
\citet{KoppShchekinov2007PhPl} with the exception of Oxygen, for which
$X_O = 8.53\cdot10^{-4}$.

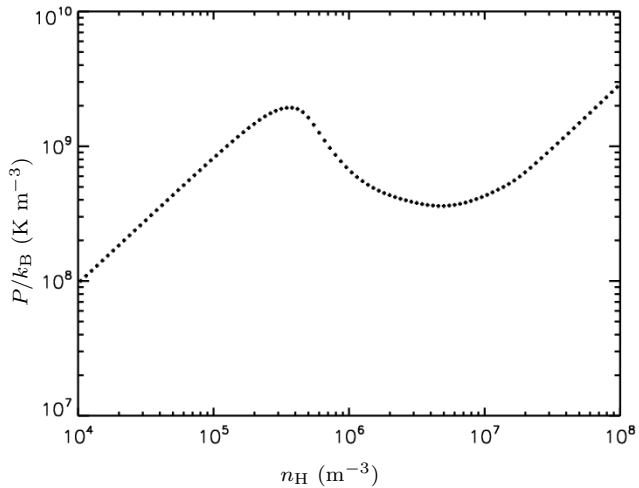
\begin{figure}
  \setlength{\unitlength}{0.08mm}
  \begin{picture}(0,0)(-50,-60)
    \includegraphics[width=1000\unitlength]{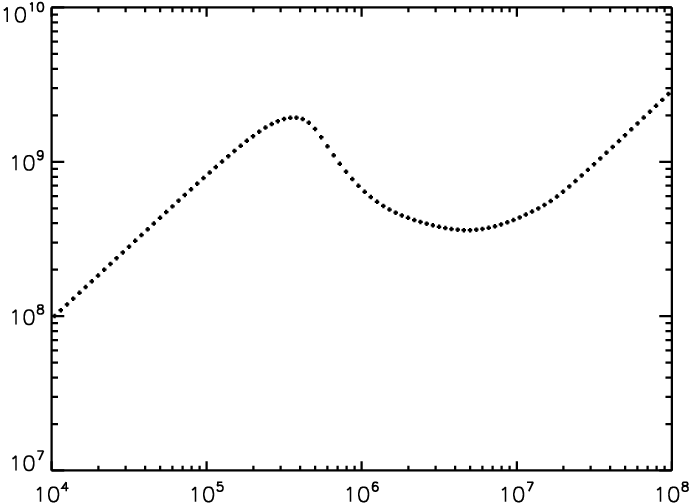}%
  \end{picture}%
  \begin{picture}(1050,797)(0,0)
    \put(10,290){\rotatebox{90}{$P/k_{\mathrm{B}}$ (K m$^{-3}$)}}%
    \put(460,0){$n_{\mathrm{H}}$ (m$^{-3})$}%
  \end{picture}%
  \caption{\label{FigPresCurve}Thermal pressure $p$ versus hydrogen
    number density $n_{\mathrm{H}}$ for the thermal equilibrium
    resulting from the heating and cooling functions used in this
    work. For details see the text.}
\end{figure}

With regard to heating processes we only take the heating by external
UV radiation incident onto the interstellar dust grains into account
\citep[see][]{BakesTielens1994}. Other heating processes are thought
to be negligible for the major part of the ISM
\citep[see][]{WolfireEtal1995}.  The energy absorbed by the dust
grains is transfered to the gas by collisions with the dust
particles. \citet{BakesTielens1994} give for the corresponding heating
rate of the interstellar gas:
\begin{equation}
  n_{\mathrm{H}} \Gamma
  =
  1.0 \cdot 10^{-25} n\ \varepsilon \ G_0
  \ \textnormal{J}\
  \textnormal{m}^{-3}\ \textnormal{s}^{-1}
\end{equation}
Here $\varepsilon$ is the fraction of FUV radiation absorbed by the dust
grains, that is converted to the heating of the gas and is mainly
determined by the neutral fraction of the gas under
consideration. This fraction is estimated by
\citeauthor{BakesTielens1994} as:
\begin{equation}
  \label{EqConvEff}
  \varepsilon = \frac{4.9 \cdot 10^{-2}}
           {1+96\left(G_0 T^{1/2}/n_{\mathrm{e}}\right)^{0.73}} +
           \frac{3.7 \cdot 10^{-2}(T/10^4 \textnormal{K})^{0.7}}
                {1+2\cdot 10^{2}\left(G_0 T^{1/2}/n_{\mathrm{e}}\right)}
\end{equation}
Here we assume that the strength of the FUV average interstellar
radiation is $G_0 = 1.7$ times the estimate by \citet{Habing1968} of
$f_{\gamma} = 1.6\cdot 10^{-6}
\textnormal{J}\ \textnormal{m}^{-2}\ \textnormal{s}^{-1}$. From the
combination of the above heating and cooling functions we computed the
thermal equilibrium curve shown in Fig. \ref{FigPresCurve}. There the
thermal pressure for thermal equilibrium is given as a function of the
hydrogen number density.

At this point, we have to keep in mind that we can only consider the
two low temperature phases of the ISM (the cold and the warm phase).
A hot phase could be described by properly incorporating ionisation of
the plasma. This, however, can not be done when using a single
magnetised fluid.

Even when using a polytropic equation of state \mbox{$p \propto
\rho^{\gamma}$} instead of the full energy equation, there arises a
slight difficulty. Namely, we have to make sure that the prescribed
kinetic energy input for the driving is in the correct ratio to the
thermal energy in the chosen phase of the ISM.

\subsection{The Driving Process}
Due to the very nature of turbulence, energy that is injected at some
scale is transformed to fluctuations of ever smaller scale, until the
dissipation range is reached, at which they are finally transformed
into heat. Thus, a process is needed which constantly replenishes
the energy that is lost at the dissipation scales, to maintain a
specific level of the turbulence. For the ISM there are several
plausible sources for this energy, of which supernovae seem to be the
most important ones.

For large-scale simulations it would be feasible to include such
driving directly into the simulations as artificial explosions
\citep[see, e. g.,][]{AvillezBreitschwerdt2004}. For local simulations
as they are performed here, however, this cannot be done. Therefore,
we include a driving process, which affects only the largest spatial
scales. On the one hand, that allows for a study of the turbulent
cascade, which is not disturbed by the driving. On the other hand,
this even allows for a plausible physical interpretation: The
fluctuations represent the cascade continuing from scales larger than
the computational domain.

The fluctuations are added as a velocity field with random amplitude
and phases. This velocity field is chosen to be incompressible in
order to mimic the transport through the inertial range from scales
larger than the numerical domain to the largest scales under
investigation. The amplitudes, however, do depend on the wave-number
$\vec{k}$, by a dependence of the variance of the corresponding
normally distributed random numbers on $\vec{k}$. By this, the average
velocity power spectrum of the driving is given by:
\begin{equation}
  E(k) = k^4 \sigma^2_{\delta \vec{u}}(k)
\end{equation}
where $\sigma_{\delta \vec{u}}(k)$ is the scale dependent standard
deviation of the amplitude of the velocity disturbance. For the
present simulations we selected a driving spectrum with $E(k) \propto
k^{-1}$. Further details of the implementation of the driving
process are given in appendix \ref{SecDrive}.

For turbulence research a high energy input of fluctuations would be
desirable. For the ISM, however, this is constrained by the actual
physical values. If, e. g., supernovae are assumed to be the sources
for the driving, there is an upper limit for the average input rate of
kinetic energy into the system. Taking into account the overall energy
output of a supernova explosion of $E \simeq 10^{43}$ -- $10^{44}$ J
\citep[see, e. g.,][]{DysonWilliams1997}, the efficiency of conversion
into kinetic energy, the volume of the galactic gaseous phase \mbox{$V_{Gas}
\simeq 1.4 \cdot10^{11}$ pc$^3$}, and the frequency of the explosions
\citep[one supernova per century -- see][]{Avillez2000}, we arrive at
a base value of about $ S_{\mathrm{e}} \simeq 10^{-27} \textnormal{ J
  m}^{-3}\textnormal{s}^{-1}$. This value may not be a strict upper
limit, but it should not be abandoned arbitrarily. The actual values
used in this work will be given when discussing the
simulations. 

\subsection{Further Issues}
One additional property, which is important for the ISM, is the
compressibility of the gas. This allows for the existence of different
kinds of shock waves. All of these have to be captured correctly by the
numerical method used for the simulations. The important issue is that
only conservative codes have been demonstrated to yield the correct
speed and structure for the different kinds of discontinuities
\citep[see, e. g.,][]{Leveque2002}. Also the steep gradients occurring
at such structures must not induce any artificial fluctuations, which
would be indiscernible from the turbulent ones.

Another issue is connected to the minimum simulation time. Classically
it might seem necessary to simulate for several sound crossing times
to allow for a signal propagation through the whole computational
domain. Here, however, we will be investigating highly supersonic
turbulence. Thus, the speed of sound is not a good tracer for the
propagation time of features in the medium anymore. Instead the
relevant timescale is the large eddy turnover time. For saturated
turbulence this is given by \citep[see][]{Frisch1995}:
\begin{equation}
  \label{EqEddyTurnover}
  \tau = \frac{L}{\vel_0}
  \quad \textnormal{with} \quad
  L=\frac{M \vel_0^{3}}{\epsilon}
  \quad \textnormal{and} \quad
  \vel_0=\left(\frac{2}{3}\frac{E}{M}\right)^{1/2}
\end{equation}
where $E$ is the overall energy and $M$ is the overall mass in the
numerical domain and $\epsilon$ is the overall dissipation
rate. The resulting values of $L$ and $\vel_0$ can be interpreted as
the typical size and the typical (rotational) velocity of an eddy,
respectively.

\section{Numerical Approach}
There are several important demands for a numerical scheme for the
simulation of the turbulent ISM. A basic requirement is a sufficiently
high Reynolds number to actually allow the system to become
turbulent. This is especially the case for the very high Reynolds
numbers present in the ISM -- corresponding to an extent of the
turbulence spectrum of some ten orders of magnitude \citep[see,
  e. g.,][]{ArmstrongEtal1995}. Even though such a simulation will not
be possible for years to come, the available simulations still have to
yield very high Reynolds numbers.

While this can -- for comparatively inefficient codes -- be remedied
by a sufficiently high spatial resolution, the high Mach number of the
plasma in some phases of the ISM is an even more important demand. The
corresponding supersonic flow in the ISM leads to the generation of
shocks \citep[see, e. g.,][]{AvillezBreitschwerdt2004}, which have to
be handled correctly by the numerical scheme. Especially, when
investigating the turbulent fluctuations there must not be any
spurious oscillations at shocks, which could not be disentangled from
the turbulence itself. Many schemes, which are able to correctly
reproduce such discontinuities are, however, of very low order, thus
gaining little from high spatial resolution.

So far, only few algorithms for numerical MHD computations are in use
for studying problems in the ISM. For two-dimensional studies see for
example \citet{PassotEtAl1995}, while for three-dimensional ones see
\citet{vestuto_etal_2003}, \citet{ChoLazarian2003},
\citet{BalsaraEtal2001}, \citet{BrandenburgDobler2002} and
\citet{MaronGoldreich2001}, the latter of which considers only an
incompressible velocity field.  Some do not really meet the above
requirements, others seem to have some additional problems \citep[see,
  e. g.,][]{Falle2002}. Here we apply our implementation of a Riemann
solver-free central scheme, which is shown to reproduce the occurring
shocks quite nicely (Essential features of this scheme are summarised
in Appendix \ref{SecDrive}). For this scheme the Reynolds number for
the ideal MHD simulations can be estimated as $Re \simeq 3\cdot 10^4$
and $Re \simeq 1\cdot 10^5$ for the simulations with 256 and 512
grid-points in each of the spatial dimensions, respectively \citep[see
also][for an estimate of the Reynolds numbers]{Kissmann2006}.

One feature of the code, which must not be forgotten, is its
conservative property. This is especially of interest for turbulence
simulations, because only using a conservative code yields the
possibility to extract meaningful information about the energy content
of the turbulence. For compressible MHD simulations in general,
however, the conservation of the quantities is even more important,
since only then a correct representation of the frequently occurring
shock waves can be assured.

\section{Simulations}
In this work we are interested in the spatial and spectral structure
of ISM turbulence. Due to the nature of the ISM it is a highly
demanding task to achieve both of these goals. Whenever the whole
structure of the ISM is considered, a global simulation would be
necessary. This, however, would not supply information about the
turbulence spectrum for individual part of the ISM. When one
concentrates, in contrast, on local patches of the ISM, it is easily
forgotten that one still has to take all the properties of the ISM
into account, which were mentioned in the introduction.

Here we present results of our numerical simulations that were
obtained for rather local models. In this work we investigate two
basic scenarios. First, we discuss our simulation results for
molecular cloud turbulence. Thereby we show the suitability of the
scheme for simulations of compressible turbulence. In this case we
investigate the influence of different equations of state on the
statistical and spatial structure of the turbulence

Second, we take the phase structure of the ISM into account. We show
local simulations for turbulence, for which the influence of heating
and cooling processes leads to a two-phase medium.

\section{Molecular Clouds}
Interstellar turbulence is often discussed in the context of molecular
clouds. This might seem arbitrary at first, but there are good reasons
to concern ourselves with these ISM structures. First, they are very
important for the dynamics of the ISM, since star formation takes
place in just these regions. Second, a far more trivial reason to
investigate turbulence in molecular clouds is that parts of the can be
assumed to be of lower complexity than other regions of the ISM. This
is due to the fact that typically the gas in molecular clouds is
shielded from the interstellar radiation field. Therefore, ionisation
and dissociation by UV photons can safely be neglected for a study of
molecular clouds. This means that also external heating is of no
importance in these regions.

The subtype of molecular cloud that we are investigating in this work
is what \citet{SnowMcCall2006} call the diffuse molecular clouds.
These are already sufficiently isolated from the interstellar
radiation field. The ionisation is, however, still sufficiently high
so that the magnetic field can strongly influence the plasma and,
therefore, must not be neglected. Despite the fact that chemistry
already plays a role in these regions it is not as complex as in the
dense cloud cores. Moreover, the number densities being of the order
of $10^8$ particles per cubic meter are still low enough that
self-gravity can be neglected. With a typical ionisation fraction of
0.01 in these molecular cloud regions, the initial electron density is
$n_{\mathrm{e}} \simeq 10^6$ m$^{-3}$. It can also be shown that
collisions of neutral and ionised particles are so frequent that a
near perfect coupling of both particle species is achieved
\citep[see][]{Kissmann2006}. Therefore, ideal MHD is still a viable
description for this environment.

The corresponding value for the temperature in these simulations is
obtained from a model for the two-phase ISM. From the discussion in
\citet{Cox2005} we find a thermal pressure of about
\mbox{$7.76\cdot10^{-14}$ Pa} meaning that we are dealing with a
temperature of about \mbox{56 K} corresponding to a sound speed of
\mbox{480 m s$^{-1}$} when we assume an average ion mass of \mbox{$m_0
  = 2m_{\mathrm{p}}$}. The latter choice reflects the fact that hydrogen occurs
partly in the form of neutral hydrogen molecules and that heavier
elements are also present in the ISM \citep[see,
  e. g.,][]{SnowMcCall2006}. For the molecular clouds we use an energy
input rate of \mbox{$1\cdot 10^{-26}$ J m$^{-3}$ s$^{-1}$}
corresponding to the fact that on average supernova explosions are
expected to occur near or even within these structures. Finally, we
choose a plasma $\beta$ of 0.3 corresponding to the observations
\citep[see, e. g.,][]{Kissmann2006}, with the initially homogeneous
magnetic field pointing into the $x$-direction.

The time is given in units of the sound crossing time through the
simulation box, which is a cube with a side-length of $L_{\mathrm{B}} =$ 40
parsecs with periodic boundaries. In these units we drive the system
with an input of kinetic fluctuations on scales $1\le kL_{\mathrm{B}}/2\pi \le
3$, where $k$ is the wave-number.  Adding the velocity fluctuations
every $\Delta t_{\mathrm{Drive}} = 10^{-3}$ we ran the simulations in
full turbulence for more than three eddy turnover times.

The results shown below were obtained with a resolution of $512$ cells
in each of the three spatial directions, corresponding to a spatial
resolution of less then 0.1 pc per cell. Here, we mainly investigate
the influence of different equations of state onto the spatial
structure and the statistics of the turbulence, where we will start
with the former of those in the next section.

\subsection{The Spatial Structure}
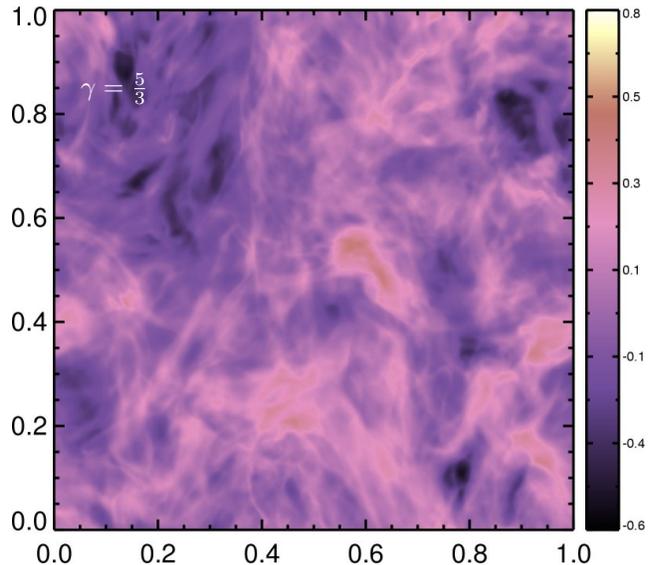
\begin{figure}
  \setlength{\unitlength}{0.084mm}
  \begin{picture}(0,0)(0,0)
    \includegraphics[width=1000\unitlength]{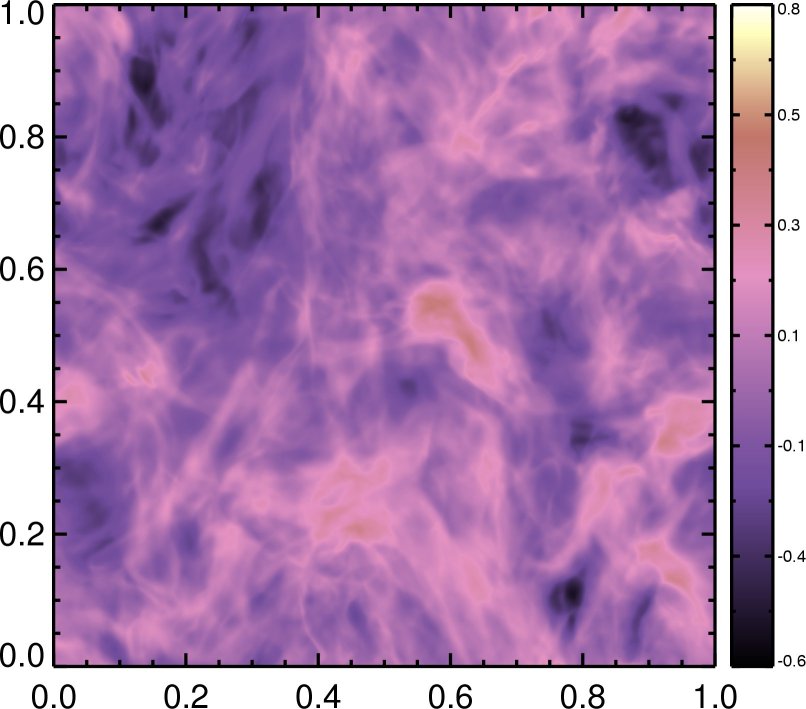}
  \end{picture}
  \begin{picture}(1000,879)(0,0)
    \put(100,740){\textcolor{white}{\large $\gamma = \frac{5}{3}$}}
  \end{picture}
  \caption{\label{FigMolDispMeasAdiab}Logarithm of the dispersion measure
    of the plasma in the numerical domain in units of $4\cdot 10^{-5}$
    pc m$^{-3}$. Results are shown for a polytropic
    index of $\gamma = 5/3$.}
\end{figure}

\begin{figure}
  \setlength{\unitlength}{0.084mm}
  \begin{picture}(0,0)(0,0)
  \includegraphics[width=1000\unitlength]{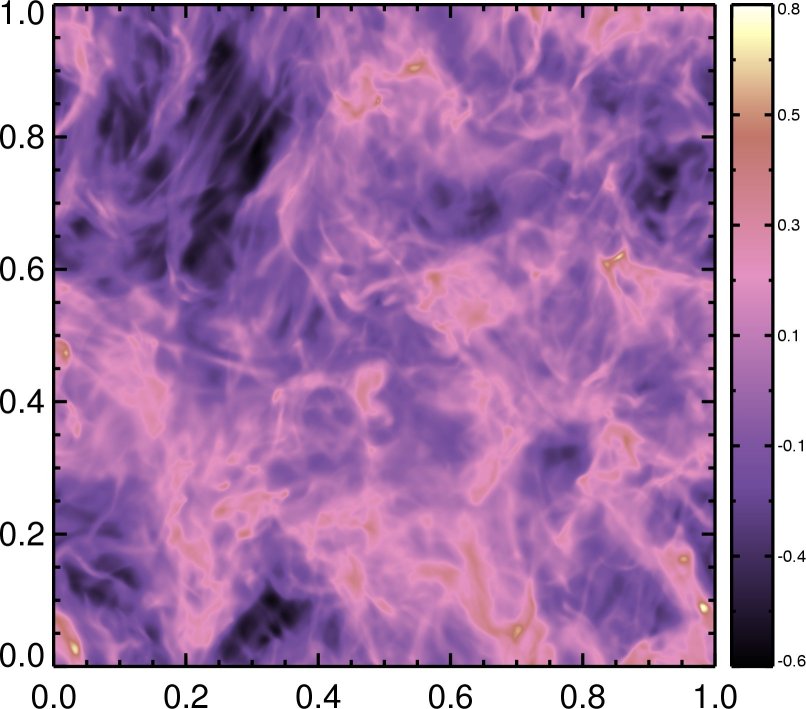}
  \end{picture}
  \begin{picture}(1000,879)(0,0)
    \put(100,740){\textcolor{white}{\large $\gamma = 1$}}
  \end{picture}
  \caption{\label{FigMolDispMeasIso} Same as Fig
    \ref{FigMolDispMeasAdiab} but for an isothermal medium.}
\end{figure}
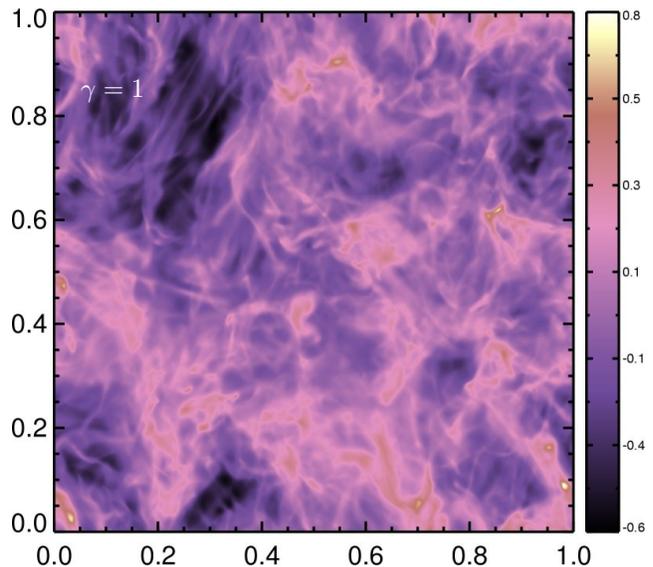

The most obvious -- and most directly accessible part of the spatial
structure -- is the density distribution. The amount of radiation
emitted by a small volume of the ISM scales in general with its
temperature and density. Therefore, for our isothermal simulations the
density is the best tracer of how the medium would look like using
different telescopes. The corresponding line-of-sight integral of the
electron density is given as:
\begin{equation}
  D_M = \int n_{\mathrm{e}} d\ l
\end{equation}
As an example we show a comparison of this dispersion measure $D_M$
for the fully turbulent state for both a polytropic and an isothermal
equation of state in Figs. \ref{FigMolDispMeasAdiab} and
\ref{FigMolDispMeasIso}.

There the results are given for an adiabatic index of $\gamma = 5/3$
in Fig.~\ref{FigMolDispMeasAdiab} and for an isothermal plasma in
Fig.~\ref{FigMolDispMeasIso}. All differences in these images can be
attributed to the equation of state, since the driving in both cases
was identical. This is reflected in the global similarities. Locally,
however, the contrast in the images for the isothermal medium is much
higher. This is due to the fact that for a polytropic equation of
state structures get smoothed due to the higher pressure.

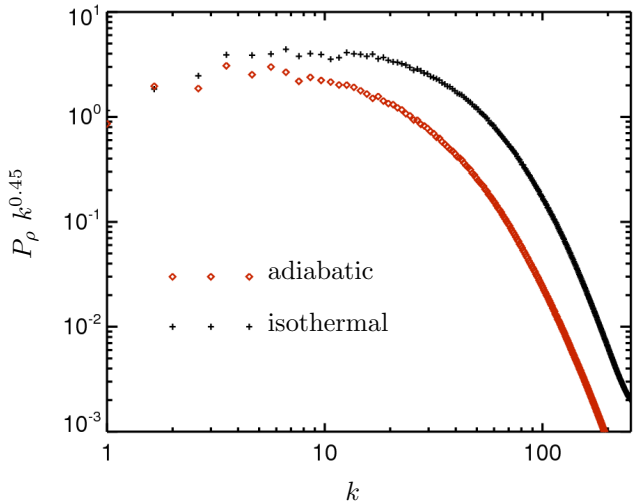
\begin{figure}
  \setlength{\unitlength}{0.07636mm}
  \begin{picture}(0,0)(-100,-65)
    \includegraphics[width=1000\unitlength]{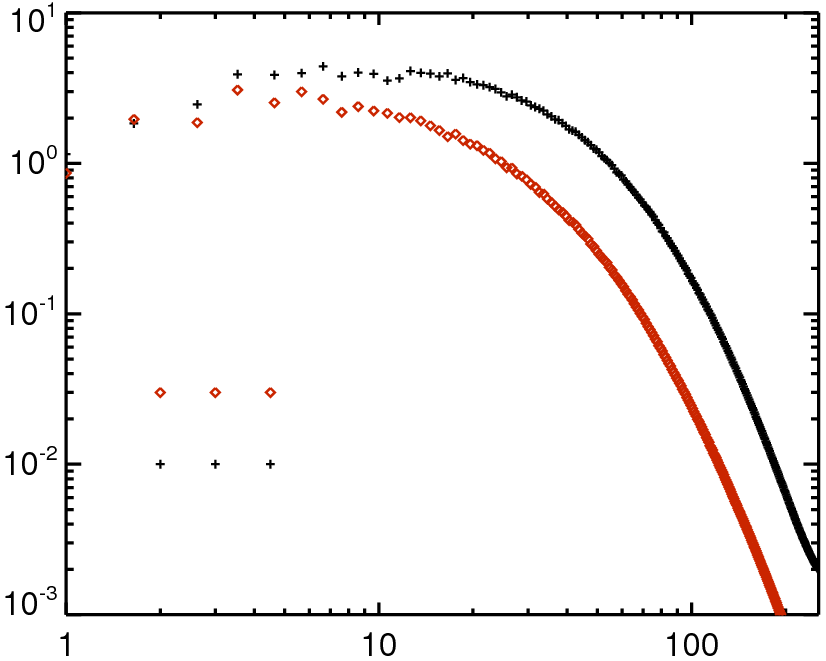}%
  \end{picture}%
  \begin{picture}(1100,885)(0,0)
    \put(595,0){\large$k$}%
    \put(10,420){\large\rotatebox{90}{$P_{\rho}\ k^{0.45}$}}%
    \put(460,295){\large isothermal}%
    \put(460,385){\large adiabatic}%
  \end{picture}%
  \caption{\label{FigMolCloudDensSpec}Density power spectra for
    molecular cloud turbulence. Here the normalised density
    fluctuation power multiplied by $k^{0.45}$ as a function of
    wave-number $k$ is shown for the adiabatic and the isothermal
    case.}
\end{figure}

This fact is also reflected in the power spectra of the density
fluctuations shown in Fig. \ref{FigMolCloudDensSpec} in a direct
comparison for both cases. For the isothermal spectrum there is
apparently much more power in the high wave-number modes of the mass
density fluctuations corresponding to the small spatial
scales. Obviously, however, not only the highest wave-numbers are
influenced by the different choices for the equation of state. The
fluctuations in the inertial range also differ.

The fact that the density fluctuation spectrum is much shallower than
the one reported in \citet{ArmstrongEtal1995} can be understood by a
selection effect regarding the observations. Other authors \citep[see,
  e. g.,][]{DeshpandeEtAl2000} report different power laws for
different phases of the ISM. It is also investigated by
\citet{KimRyu2005}, how this power law index depends on the Mach
number of the medium. Our results shown in
Fig. \ref{FigMolCloudDensSpec} are consistent with these latter
results, where the authors also find a small power law index for high
Mach numbers.

One important aspect of the spatial structure is the distribution of
shocks because they are thought to have a direct influence on the
statistics of the turbulence. This is, e. g., reflected in the model
for the structure functions by \citet{SheLeveque1994}. This model
indicates that the exponents of the structure functions -- and, thus,
the index of the power law of the velocity power spectrum -- depend on
the dimensionality of the dissipative structures.

\begin{figure}
  \resizebox{84mm}{!}{
    \rotatebox{270}{
      \includegraphics[width=\textwidth]{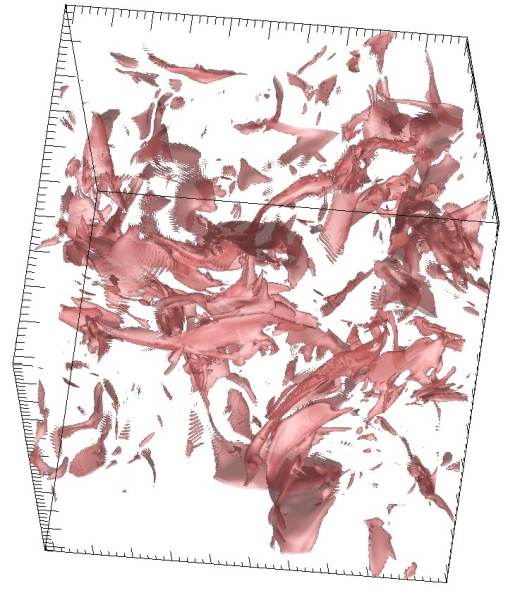}
    }
  }
  \caption{\label{FigMolCloudCurrDensIso}Isosurface plots for the
    square of the normalised current density. Results are shown for
    the isothermal case.}
\end{figure}
As an example we show in Fig. \ref{FigMolCloudCurrDensIso} the
distribution of current sheets for the isothermal medium. Even though
this plot differs from that for the adiabatic case (not shown), the
essential information remains identical. The dissipative structures
are current \emph{sheets}. Thus, the fluctuation spectrum is dominated
by two-dimensional structures in both cases. This leads, according to
the model by \citet{SheLeveque1994}, to the expectation that the
spectrum and also the structure functions should by identical to the
bounds of numerical accuracy.

\subsection{Statistics}
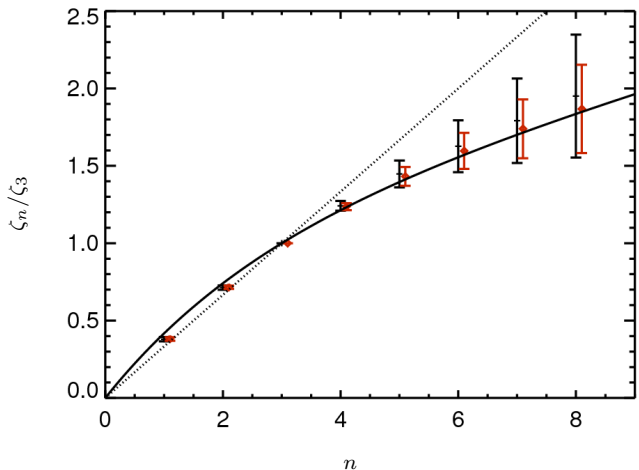
\begin{figure}
  \setlength{\unitlength}{0.07636mm}
  \begin{picture}(0,0)(-100,-65)
    \includegraphics[width=1000\unitlength]{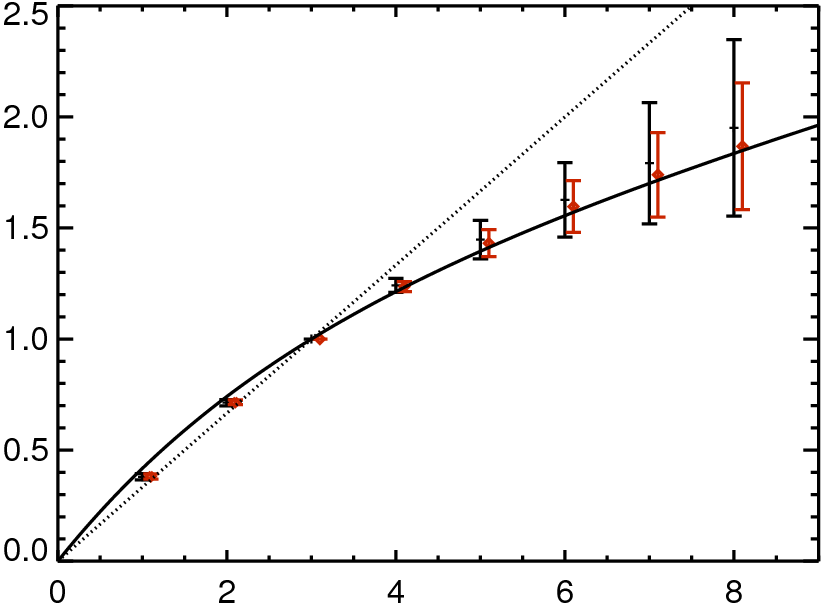}%
  \end{picture}%
  \begin{picture}(1100,803)(0,0)
    \put(585,0){$n$}%
    \put(0,415){\rotatebox{90}{$\zeta_n/\zeta_3$}}
  \end{picture}%
  \caption{\label{FigMolCloudVeloStruct}Exponents for the structure
    functions as a function of order $n$ obtained using the extended
    self-similarity. Here data points for the isothermal medium are
    given in black, whereas the results for the adiabatic medium are
    shown in red. The latter are shifted a little to the right to
    enhance visibility. The data are given together with the
    corresponding error bars. Additionally, we show the theoretical
    prediction by \citet{BoldyrevEtal2002} as the solid line and the
    \citet{Kolmogorov1941c} as the dotted line.}
\end{figure}
The statistics of the turbulence is shown here in the form of the
velocity structure functions in Fig. \ref{FigMolCloudVeloStruct}. Here
we investigate the parallel structure functions $S_n(l)$:
\begin{equation}
  S_n(l) = \left<\left|\left(
  \vec{v}(\vec{r} + \vec{l}) - \vec{v}(\vec{r})
  \right)\cdot \vec{e}_l\right|^n\right> \propto l^{\zeta_n}
\end{equation}
with $\vec{e}_l = \vec{l}/l$ and $<>$ indicates an averaging over the
spatial domain for one time-step. In Fig. \ref{FigMolCloudVeloStruct}
we give the exponents of these structure functions as a function of
their order $n$. These were computed from a single time-step using the
extended self-similarity \citep[see][]{BenziEtAl1993}. The latter must
be invoked here, since the inertial range is not visible for the
structure functions themselves. The statistics is quite good for the
lower order structure functions. As can be seen from
Fig. \ref{FigMolCloudVeloStruct} the results are in good agreements to
the theoretical prediction given in \citet{BoldyrevEtal2002} for
compressible MHD:
\begin{equation}
  \frac{\zeta_n}{\zeta_3} = \frac{n}{9} + 1 -
  \left(\frac{1}{3}\right)^{n/3}
\end{equation}
In particular it is obvious that the data do contradict the classical
Kolmogorov scaling of $\zeta_n/\zeta_3 = n/3$. Therefore, the results
confirm the \citeauthor{SheLeveque1994} model. For both the isothermal
and the adiabatic medium the spectral slope in the inertial range can
be computed from the second-order structure function to be about
$1.71$, which is just a little steeper than the classical Kolmogorov
power law index.

Obviously there is no principal difference between the turbulence
statistics of the isothermal and the adiabatic simulations. The
structure functions are very similar for both cases. This is further
evidence of the basic scaling structure. The shock structure might be
different for isothermal or adiabatic systems, but the essential
aspect of the \citeauthor{SheLeveque1994} model is the dimensionality
of the dissipative structures. This is especially the case in ideal
MHD, where shocks would have no extent in the shock direction if no
numerical dissipation were present.

Concluding, we can say that the statistics of the turbulence does not
depend on the equation of state used for the computations. There is,
however, a strong dependence of the spatial structure as was seen in
the preceding paragraph. Therefore, the equation of state has to be
considered carefully whenever any resemblance to actual observations
of molecular clouds is sought.

\section{Diffuse Interstellar Gas}
For the ISM, turbulence is mostly discussed for the context of
molecular clouds. There are, however, also more global simulations for
the interstellar medium of the Milky Way \citep[see,
  e. g.,][]{AvillezBreitschwerdt2004,BreitschwerdtAvillez2006}. These
aim at a spatial domain reaching from the Galactic plane up into the
halo of the Galaxy in order to investigate, e. g., the stratification
of the interstellar medium.  Here we are interested in similar
simulations, however, on a local scale. In this paragraph, we will
give our first results on such simulations, thus, showing the
feasibility of the code for such simulations.

As was already mentioned above, the most important difference between
molecular gas and the diffuse phase of the ISM is that the latter is
influenced by the interstellar radiation field. This will be taken
into account by the inclusion of an appropriate heating and cooling
function for the diffuse ISM. Therefore, we solve the full energy
equation with the additional source terms supplied by heating an
cooling. In this case, in contrast to the simulations for the
molecular clouds, we do not have to prescribe the correct temperature
of the gas -- this is instead achieved by heating and cooling
processes. We have, however, to be careful to choose an initial
condition from which the medium can evolve into a state possibly
realised in nature. The hope that the initial state will independently
evolve to a physically correct state is ruined, unfortunately, by the
periodic boundary conditions. Due to these it is, e. g., not possible
for high pressure regions to push anything out of the numerical domain
-- that is, the mass in the numerical domain will stay constant
throughout the whole simulation. Therefore, the average pressure is
just determined by the temperature.

This temperature will, at least on average, not exceed the value
$10^{4}$ K due to the strongly increasing line cooling efficiency
above this temperature.  Thus, the initial density must be chosen
sufficiently high, so that we end up with the correct pressure for the
ISM. Due to the inclusion of the cooling function we will eventually
end up with a dynamic equilibrium of a two-phase medium inherent in
the cooling function. The initially homogeneous medium will be
compressed into a dense, cold phase with a warm phase filling the
space in between. The space filling factors of these phases will most
probably depend on the initial density, with a lower density leading
to a dominance of the warm phase, which we are essentially looking
for.

By several numerical tests we arrived at a suitable initial density of
$n_0 = 2\cdot10^6$ m$^{-3}$, yielding physically reasonable results
for the pressure distribution. Keeping in mind that the typical
density of the warm \textup{H\,{\mdseries\textsc{i}}} gas is about
$10^5$ m$^{-3}$, this allows for a fragmentation into a warm
\textup{H\,{\mdseries\textsc{i}}} phase and a cold phase. The
temperature was then initialised by the equilibrium value resulting
from the heating-cooling equilibrium. With a side length of the
numerical domain of 40 pc we apply an energy input rate of
\mbox{$S_{\mathrm{e}0} = 5.3\cdot10^{-27}$ J m$^{-3}$ s$^{-1}$}. This
rate was deliberately chosen to be moderately lower than the rate used
for the molecular cloud simulations, because for those we expect the
energy input into the turbulence to be higher than for the diffuse
interstellar gas (DIG) due to the immediate presence of the sources of
kinetic energy. Finally, the initial magnetic induction is set to
about the same value as it was used for the molecular clouds.

The driving was implemented in the same way as for the molecular
clouds. Here the simulations were performed on a grid of 256 cells in
each of the spatial dimensions (due to the much higher numerical cost
as compared to the molecular cloud simulations -- this is mainly due
to the sharp density contrast effected by the thermal instability,
which results in a smaller time-step. Apart from that we had to let
the simulations run for a longer time before a quasi-steady state was
reached.). They were again evolved for several eddy turnover times (in
this case we integrated up to 1.5 sound crossing times) after a
saturated turbulent state was reached.

After the turbulence is fully developed the energy injected to
maintain the turbulence is converted into heat at the small spatial
scales, thus yielding a spatially non-homogeneous heating
process. Thereby the average temperature of the medium increases to a
value higher than the equilibrium temperature computed from the
heating and cooling functions.

\subsection{The Two-Phase Medium}
As was expected our simulations indeed yielded a two-phase medium.
Here the resulting temperature for the warm phase of the diffuse
interstellar gas is of special interest due to the ongoing discussion
about the heating processes for this phase of the ISM. As is
illustrated in \citet{Reynolds1995} and \citet{ReynoldsEtAl1999} the
heating process for the dilute ionised gas known as the \emph{diffuse
  interstellar gas} or the \emph{warm ionised medium} is not entirely
clear yet. A possible solution of this problem was suggested in
\citet{MinterSpangler1997} and was extended by
\citet{SpanierSchlickeiser2005}. The authors suggest the decay of
interstellar plasma turbulence to be the main heating source for this
environment.

The main shortcoming of these theoretical models is the fact that the
authors had to use an analytical model for the turbulence
spectrum. Therefore, the dissipation of the turbulence and the
replenishing of the fluctuation energy in the energy range do not
happen self-consistently. This could easily lead to incorrect results
for the actual heating rate, when there is, e. g., too much energy in
the model for the dissipation range. This is surely not the case for
our numerical simulations. The results presented have, nonetheless, to
be seen as a preliminary test for the study of the heating of the
DIG. This is due to the fact that we use the well-documented cooling
function for the warm \textup{H\,{\mdseries\textsc{i}}} gas, with a
fixed degree of ionisation of 0.1, whereas major parts of the DIG have
rather to be regarded as warm
\textup{H\,{\mdseries\textsc{ii}}}. There is as yet, however, no
analytical model available for the cooling of the warm
\textup{H\,{\mdseries\textsc{ii}}} with a varying degree of ionisation
-- especially when using the single-fluid MHD equations.  The best
choice would, obviously, be the use of a multi-fluid model where
ionisation and recombination are included, but such a model can not be
solved using current numerical methods for a highly compressible
medium.

\begin{figure}
  \includegraphics[width=84mm]{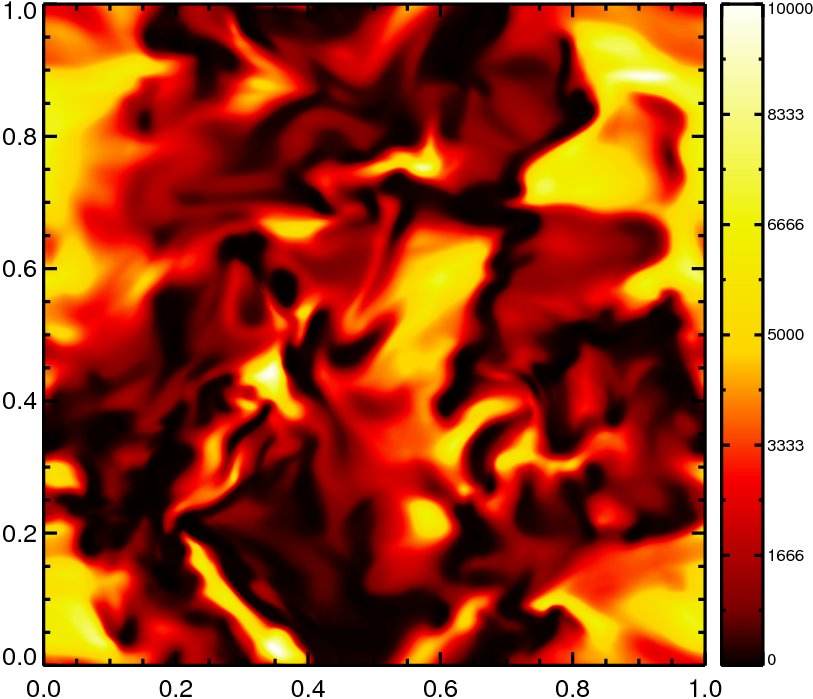}
  \caption{\label{FigDIGTempCut} The temperature of the plasma in a
    slice through the computational domain given in Kelvins according
    to the colour-coding. Here the scale of the box is 40 pc.}
\end{figure}
\begin{figure}
  \setlength{\unitlength}{0.07636mm}
  \begin{picture}(1100,733)(-100,-100) 
    \put(365,-70){$\log(n [\textnormal{m}^{-3}])$}%
    \put(-70,200){\rotatebox{90}{$\log(e_{th}/k_{\textnormal{B}}[\textnormal{K m}^{-3}])$}}%
    \includegraphics[width=1000\unitlength]{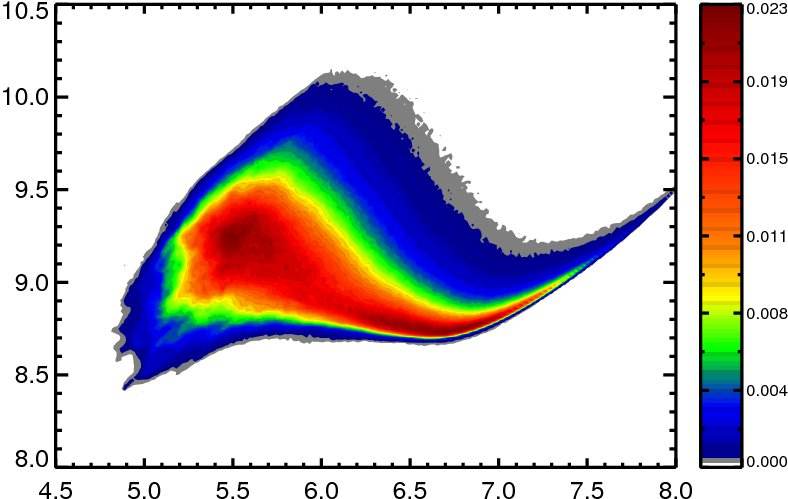}%
  \end{picture}
  \caption{\label{FigDIGTempDist} Phase space distribution function
    for the whole computational domain. Obviously the unstable regime
    between the warm ($n < 5.6$ m$^{-3}$) and the cold phase ($n >
    6.8$ m$^{-3}$) is significantly populated due to the turbulence.}
\end{figure}
The spatial temperature distribution as found from our simulations is
visualised in Fig. \ref{FigDIGTempCut}. The computed ISM is dominated
by a warm phase with temperatures of several thousand Kelvins. Regions
containing this warm gas are separated by cool clouds with
temperatures below \mbox{500 K}. These can be identified as the
typical molecular cloud structures. Thus, we have reproduced what is
commonly known from observations -- cool clouds embedded within a warm
inter-cloud medium.

With regard to Fig. \ref{FigDIGTempCut} there is apparently a certain
spatial scale connected to the fluctuation in the numerical
domain. This scale comes about by the thermal instability. It arises
due to the fact that fluctuations with sufficiently short wavelength
are stable against the thermal instability when their period is
shorter than their cooling time, whereas fluctuations with long
wavelength are unstable. Thus, there is an intermediate scale, for
which the growth rate for the thermal instability is maximal. This
scale can be estimated by the product of the cooling time with the
typical wave speed. From the fast-mode speed we, thus, find a scale of
$\sim$6.5 pc for the maximally growing modes, which is consistent with
the impression from Fig. \ref{FigDIGTempCut}.

For a more direct analysis of the present phase structure we show the
phase-space distribution in Fig. \ref{FigDIGTempDist}. Apparently the
thermal equilibrium curve leaves its footprint in this
distribution. While a warm phase with densities below $n \simeq 5.6$
m$^{-3}$ and a cool phase with densities above $n \simeq 6.8$ m$^{-3}$
are present it is, however, obvious that the unstable range is also
significantly populated. This is due to the fact that the gas is
driven into the unstable regime by the turbulence. For further
discussions on the population of the unstable regime see also
\citet{Sachnchez-SalcedoEtAl2002ApJ}

\subsection{Polarisation Canals}
Apart from the two-phase structure, which is directly accessible to
observations, there are also other quantities available to
observers. One way to gain knowledge about the interstellar magnetic
field is connected to the Faraday effect. The observational quantity
in this context is the so-called rotation measure $R_M$. This can be
accessed observationally, e. g., by an investigation of the polarised
pulsar radiation. For simulations, this quantity can be accessed
directly by integrating the data over the numerical domain.

Instead of the rotation measure, however, we will rather discuss the
polarised intensity, which is closely connected to the former. We
investigate a phenomenon found in several radio maps: elongated
structures with very little polarised intensity with a width near that
of the telescope beam -- being usually referred to as canals
\citep[see, e. g.,][]{FletcherShukurov2006}. Basically they are
thought to result at least partially from the finite resolution of the
corresponding radio observations. Whenever the polarisation of the
detected radiation changes over an area on the sky smaller than the
beam of the telescope, these polarisation canals can occur.

Here we will investigate the possibility of the formation of these
canals due to a foreground polarisation screen as suggested in
\citet{FletcherShukurov2006}. The other possible method for the
formation of such canals discussed there -- a formation due to
differential Faraday rotation -- is left for the future. Thus, we
assume that homogeneously polarised radiation is produced in a region
behind a foreground medium and is subject to Faraday rotation only in
the latter. Strong gradients in this foreground Faraday screen result
also in strong gradients in the Faraday rotation for radiation from
the background. This then leads to the observed canals, whenever
radiation of perpendicular polarisation occurs over the width of a
telescope beam.

In this case the pattern of the observed canals strongly depends on
the beam-width and also on the wavelength of the observed
radiation. For the latter a too short wavelength means that the
foreground Faraday rotation is too ineffective as to produce different
polarisation angles. In contrast a too long wavelength causes complete
depolarisation -- even a slight difference in the rotation measure
causes a huge difference in polarisation, so that the polarisation
vectors become randomly distributed.

\begin{figure}
  \resizebox{84mm}{!}{
    \includegraphics{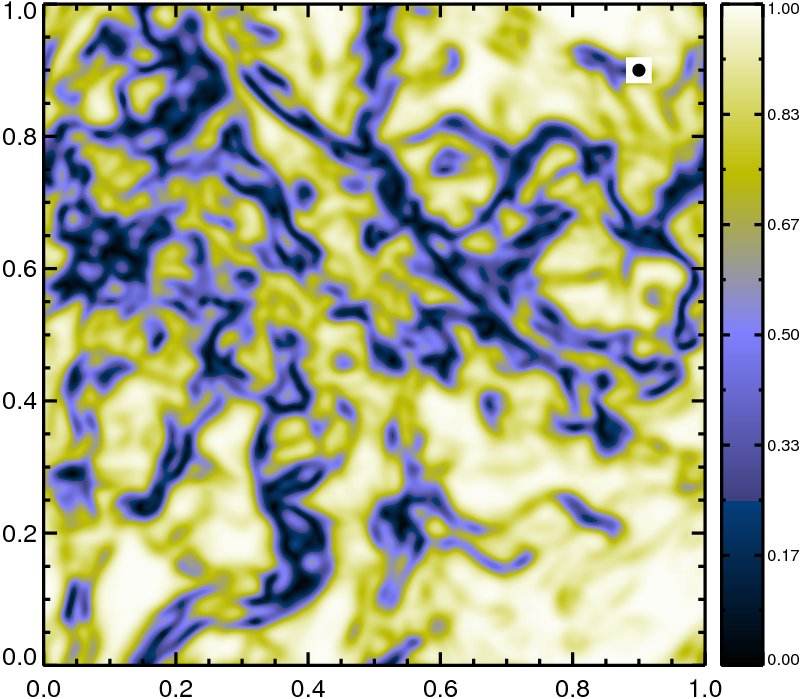}
  }
  \caption{\label{FigDIGPolInt}Normalised polarised intensity computed
    assuming a homogeneously polarised background radiation. The
    resulting intensity is computed with the plasma in the simulations
    being used as a foreground Faraday screen. The line-of-sight
    corresponds to the initial magnetic field direction. Results are
    shown for a wavelength, $\lambda$, of 0.3 m. The width of the
    artificial telescope beam is indicated as the black circle in the
    white box.}
\end{figure}

We computed the polarisation maps corresponding to our simulations
using a similar procedure as introduced in
\citet{HaverkornHeitsch2004}. From the rotation measure for the
corresponding direction we first compute the local polarisation angle
$\delta \theta$, which is given by:
\begin{align}
  \label{EqFaradayRot}
  \delta \theta(\lambda) &= R_M \lambda^2
  \\
  \qquad &\textnormal{with} \qquad
  R_M =
  - \frac{e^3}{8 \pi^2\varepsilon_0 c^3 m^2_e}
  \int_0^l n_{\mathrm{e}}(l') B_{\parallel}(l')\ dl' 
  \nonumber
\end{align}
From this we then compute the Stokes parameters $Q$ and $U$:
\begin{equation}
  U = \sin (2 \delta\ \theta(\lambda))
  \quad
  \textnormal{and}
  \quad
  Q = \cos (2 \delta\ \theta(\lambda))
\end{equation}
By smoothing their maps using a Gaussian beam with a width of 2.5 grid
cells we then mimic the beam width of the virtual telescope. From
these smoothed maps $\tilde U$ and $\tilde Q$ we finally compute the
polarised intensity:
\begin{equation}
  I = \sqrt{\tilde U^2 + \tilde Q^2}
\end{equation}
Be aware that the results shown in Fig. \ref{FigDIGPolInt} were
computed for a medium in a box of an extent of only 40 parsecs. Unlike
\citet{HaverkornHeitsch2004} we did not stack several of the
simulation boxes atop of each other in the direction under
consideration with a shift in a perpendicular direction. We rather
intend to use an adapted wavelength that yields canals in the
polarised intensity. Whereas the wavelength used might not correspond
to any of those used for actual observations, an artificial extension
of the numerical domain would just yield similar results for a shorter
wavelength. We, however, do not feel confident stacking shifted boxes
atop of each other, because the shifts of the individual boxes
against each other can lead to unphysical gradients at their
interfaces.

The polarised intensity map computed from our simulations is depicted
in Fig. \ref{FigDIGPolInt} for the directions along the initial
magnetic field. We show the resulting intensity for a wavelength of
$\lambda = 0.3$ m best suited to visualise the polarisation
canals. Interestingly, we had to choose a slightly different
wavelength ($\lambda = 0.4$ m) for the direction perpendicular to the
magnetic field (not shown here). Obviously the gradients of the
rotation measure in the direction of the initial magnetic field are
stronger than the ones in the perpendicular direction. Therefore, the
wavelength for this case had to be chosen a little shorter than for
the other one.  Apparently, it is possible to explain the
depolarisation canals visible in radio observations by a foreground
Faraday screen. There are obvious canals of just the width of the
telescope beam. One has, however, to be careful with this statement:
we found these canals only to become apparent in a very limited range
of wavelengths. For actual observations it is highly improbable to
observe just the appropriate wavelength range. Nonetheless, we see an
encouraging similarity when we compare our results to actual
observations \citep[see, e. g.,][]{ShukurovBerkhuijsen2003}. Thus, it
might be that the observed effect is due to the quite similar effect
of the differential Faraday rotation.

Our results also look quite similar to those obtained by
\citet{HaverkornHeitsch2004}. In contrast to their approach we used a
physically correct representation of the warm phase of the
ISM. Especially the Mach number chosen here is far more appropriate
for the dilute plasma than the one used by these authors. Their
statement that the turbulence cascade is very similar for the high
Mach number and the low Mach number regime has to be considered with
caution. One important point is that the spatial density structure has
a huge influence on the rotation measure as will become clear
later-on. Also, as was discussed in \citet{KimRyu2005} the density
structure is very different for different Mach numbers. Consequently,
one has to be very careful when using high Mach number simulations for
the dilute ISM.

\begin{figure*}
  \centering
  \includegraphics[width=8.5cm]{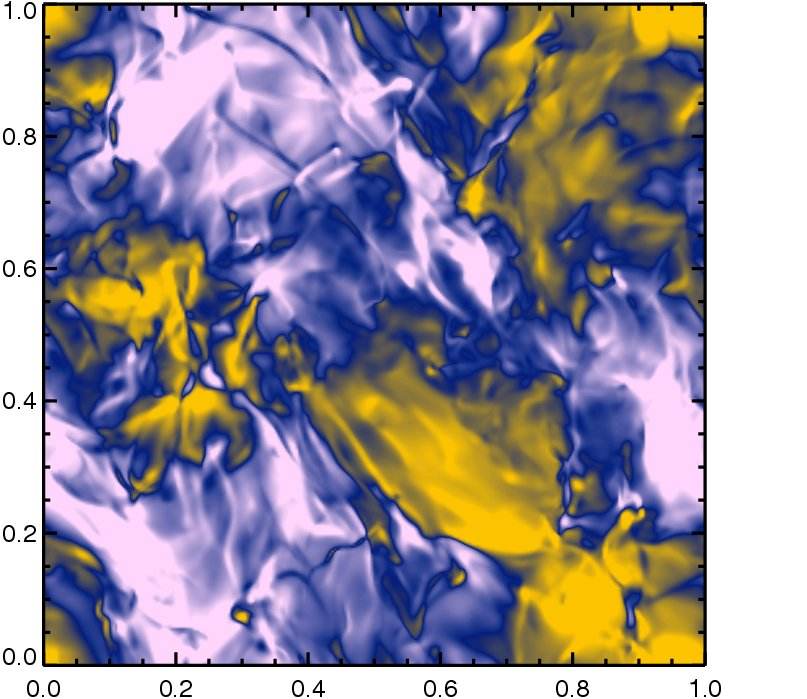}
  \hfill
  \includegraphics[width=8.5cm]{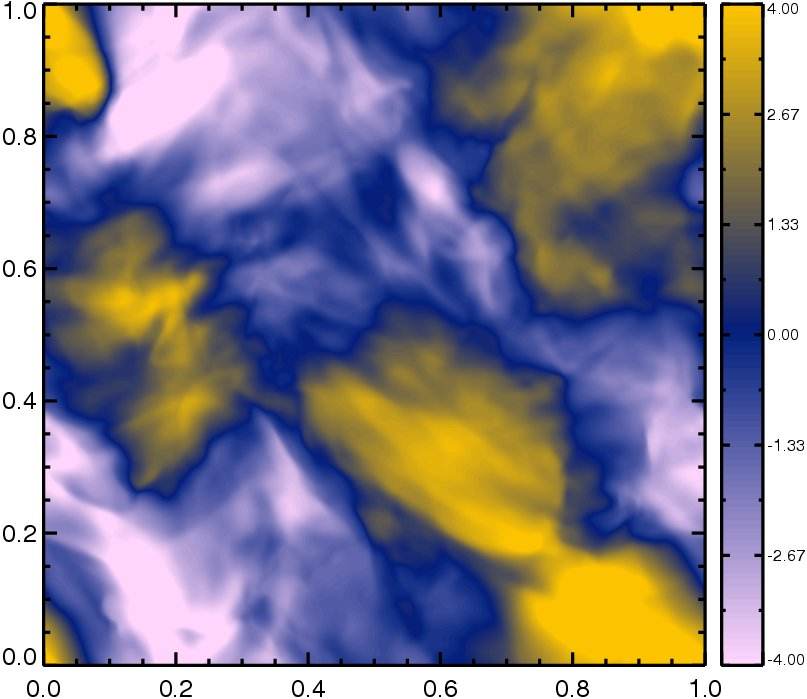}
  \caption{\label{FigWIMMagVergl}Comparison of the mass-weighted
    average for the normalised parallel magnetic induction (left) to
    the actual line-of-sight integral (right).}
\end{figure*}
The latter statement is supported by what is displayed in
Fig. \ref{FigWIMMagVergl}, where the magnetic induction as inferred
from the rotation and the dispersion measure is compared with the
average magnetic induction along the same direction. Clearly the
estimate differs significantly from the actual values - this shows the
strong difference between the mass-weighted and the unweighted
average. In particular we find that the mass-weighted average shows
much more structure than is actually there. What is additionally
apparent in Fig. \ref{FigWIMMagVergl} is that the absolute value of
the magnetic induction is on average slightly overestimated as
compared to the actual line-of-sight integral. This qualitative
impression is confirmed by a quantitative analysis -- on average the
magnitude of the magnetic induction turns out to be higher by a factor
of up to 1.15. This has to be taken into account whenever an
interpretation of any observations is attempted. The reason for this
discrepancy is the inhomogeneous mass distribution. For a completely
homogeneous medium the rotation measure would directly yield the
average magnetic induction. Obviously, the Mach number for the dilute
phase of the ISM is still high enough to yield this strong discrepancy
between the actual and the inferred magnetic field.

This result is another hint that one has to be careful to use the
appropriate Mach numbers for the simulations of all the ISM phases. It
also applies to the computation of the polarisation canals. Due to the
fact that the rotation measure seems to depend strongly on the Mach
number, one has to take care to use the appropriate simulations for
the polarisation canals. It will be left as a task for the future to
investigate how the deviation of the estimated magnetic field strength
and also the structure of the canals depend on the average Mach number
of the medium.

\section{Conclusions}
In this work we investigated basic turbulence theory in the framework
of the interstellar medium. In many cases turbulence simulations are
applied to the interstellar medium (ISM) merely because it is a medium
with extremely high Reynolds numbers, and the parameters of the ISM
are only taken into account as far as they are needed for the
turbulence research. Here, however, we investigated basic turbulence
properties, while at the same time we modelled the properties of the
ISM as thoroughly as possible. The important point is that there are
many physical processes operating in the ISM, which eventually should
be incorporated in the corresponding simulations. These processes
reach from external influences of the radiation field originating from
hot stars to the internal interaction of the particles culminating in
the intricate chemistry of the molecular cloud medium. Each of the
different phases of the ISM has its own dominant processes to be taken
into account for a realistic modelling.

In the present simulations we included heating and cooling where is
was necessary. The resulting equations were then solved using a
Riemann solver-free conservative finite-volume scheme. From the
simulation results it becomes apparent that this scheme is suitable
for the simulation of ISM turbulence. In this paper we investigated
two different scenarios for the ISM turbulence. We simulated both an
isothermal molecular cloud medium and the dilute phase of the ISM.

For the former we demonstrated that the turbulence in our molecular
cloud simulations is in accord with most recent models for
compressible MHD turbulence. We find the dissipative structures in our
simulations to be mainly two-dimensional. Shocks and current sheets
make the major contribution to the damping of the turbulent
fluctuations. From the computed second-order structure function we can
derive the power law index for the power spectrum of the velocity
fluctuations. This is only slightly steeper than that for
incompressible, homogeneous hydrodynamical turbulence, i. e., it
remains considerably below 2 for all cases. Therefore, this exponent
is comfortably far away from what was classically thought to be a
special value for particle transport theories \citep[see,
  e. g.,][]{Schlickeiser1988}.

Further, we investigated the consequences of using an isothermal and
an adiabatic equation of state. While there are clear differences in
the spatial structures, these do not extend to the statistics of the
turbulence that can be characterised by structure functions. Only for
the density power spectrum there are significant differences, where,
however, simulations with still higher spatial resolutions are needed
to clarify more details. It might be possible that these differences
result from the smoother density structures of the adiabatic medium,
which would sharpen for higher spatial resolutions.

Also for the warm phase of the ISM we could demonstrate that the
numerical scheme is appropriate for simulations of this phase.
The observed temperatures were found to be in the range known for the
DIG. The warm gas surrounds in these simulations dense filaments of a
cold gas, which corresponds to the typical molecular cloud
parameters. This two-phase medium is, however, not in thermal pressure
equilibrium. The ISM obviously has to be regarded as a very dynamic
medium.

Finally, we discussed the rotation measure obtained from the numerical
simulations. We could identify canals also seen in polarised intensity
maps of the ISM found from radio observations. In this context we
showed that it is important to use the correct Mach number in the
simulations. Even for comparatively low Mach numbers for the WIM there
can obviously result a significant deviation of the estimate of the
magnetic field inferred from the observation of the rotation measure
as compared to the actual magnetic field.

\subsection*{Acknowledgements}
We would like to thank the anonymous referee for a timely and useful
report. This work was partially financed by the Deutsche
Forschungsgemeinschaft (DFG) through the Sonderforschungsbereich SFB
591 and benefitted from the Finnish-German cooperation funded via DAAD
313-SF-PPP Finnland. Computational resources were provided in the
project hbo25 by the Forschungszentrum J\"ulich.

\appendix
\section{The Numerical Solver}
\label{SecCweno}
The hyperbolic part of the system of equations introduced in section
\ref{equations} is solved using a multi-dimensional version of the
semi-discrete central scheme introduced in
\citet{KurganovNoellePetrova2001}. This scheme was designed to yield
highly accurate solutions of hyperbolic conservation laws of the
general form:
\begin{equation}
\label{evolution}
\frac{\partial \mathbf{u}}{\partial t} + \nabla \cdot \tens{F}(\mathbf{u}) = 0
\end{equation}
Here $\mathbf{u}$ is some vector quantity with $\tens{F}$ being a
Flux-Tensor, which may also depend on $\mathbf{u}$. The resulting
scheme is derived using an integration over \emph{parts} of the
individual numerical cells in order to obtain a numerical flux over
the cell interfaces without the necessity to employ a Riemann
solver. This together with a TVD (total variation diminishing)
reconstruction of the point values at the cell interfaces from the
resulting cell averages, yields a simple and robust numerical scheme,
which suppresses spurious oscillations at steep gradients.

In the limit $t \to 0$ one approaches the so-called
\emph{semi-discrete} version of the scheme. In this limit the scheme
can be written in the general form:
\begin{align}
  \label{EqCweno}
  \frac{\partial \vec{u}_{i,j,k}}{\partial t} =
  &\frac{\vec{H}^x_{i+1/2,j,k} - \vec{H}^x_{i-1/2,j,k}}{\Delta x} +
  \frac{\vec{H}^y_{i,j+1/2,k} - \vec{H}^y_{i,j-1/2,k}}{\Delta y} 
  \nonumber\\
  +&\frac{\vec{H}^z_{i,j,k+1/2} - \vec{H}^z_{i,j,k-1/2}}{\Delta z}
\end{align}
where the fluxes are given by:
\begin{align}
  \vec{H}^x_{i+\half,j,k} 
  =&
  \frac{1}{a^+_{i+\half,j,k} + a^-_{i+\half,j,k}}
  \\
  &\times\left(a^+_{i+\half,j,k} \vec{f}(\vec{u}^-_{i+\half,j,k}) -
  a^-_{i+\half,j,k} \vec{f}(\vec{u}^+_{i+\half,j,k})\right.
  \nonumber\\
  &\quad- \left.a^+_{i+\half,j,k} a^-_{i+\half,j,k}
  \left(\vec{u}^+_{i+\half,j,k} - \vec{u}^-_{i+\half,j,k}\right)
  \right)
  \nonumber
\end{align}
and similarly for $H^y$ and $H^z$. Here $a^{\pm}$ represent an upper
limit for the maximum signal propagation velocity, $\vec{u}^{\pm}$
indicates the point values obtained using the reconstruction
polynomial in the local ($-$) and in the next cell ($+$),
respectively, and $\vec{f}$ denotes the corresponding physical fluxes.
In the case of the MHD equations the maximum signal propagation
velocity can be estimated to be the sum of the background flow
velocity and the fast magneto-sonic speed.

To ensure the TVD property of the resulting scheme the point values
$\vec{u}^{E/W}$ at the cell faces have to be estimated in a way as to
be TVD. This is done via a reconstruction polynomial the order of
which at the same time determines the order of the scheme. This
reconstruction polynomial is chosen in a way as to fulfil the
conservation properties -- in particular we are using a second-order
polynomial together with a minmod limiter \citep[for a proof of the
  TVD property and a discussion of the minmod limiter see,
  e. g.,][]{KurganovNoellePetrova2001}.
\\
\underline{Remarks}:
\begin{description}
\item 
  The order of the scheme is not only given by the order of the
  reconstruction polynomial. Also, the numerical approximation of the
  integral over the cell faces has to be at least of the same
  order. This was an additional motivation to use a second-order
  scheme, because for second-order the midpoint rule is still
  sufficient. For higher-order schemes the reconstruction, therefore,
  becomes much more expensive. Apart from that the order has to be
  reduced for higher-order schemes near any of the omnipresent shock
  waves in highly supersonic turbulence.
\item
  So far, we only have a scheme capable of solving hyperbolic
  equations. As we will see in the next section, the solenoidality of
  the magnetic field demands some additional effort.
\item
  The time evolution of the system of MHD equations is done via a
  Runge-Kutta method.
\end{description}

\subsection{Divergence of the Magnetic Field}\label{DivClean}
Whereas the analytical derivation of the MHD equations did consider
the solenoidality of the magnetic field by application of the
constraint:
\begin{equation}
  \nabla \cdot \vec{B} = 0
\end{equation}
Eq. (\ref{EqInduction}) does not maintain this constraint
inherently. This leads to the problem that $|\nabla \cdot \vec{B}|$
increases with time, thus, rendering the solution unphysical.

Luckily, there are different ways to fulfil this condition
numerically. Unfortunately only one of these could be shown to be
suitable for the simulation of compressible turbulence. Especially the
\emph{projection scheme} proposed first by \citet{Chorin1968} for the
incompressibility constraint in the Navier-Stokes equations and
later-on applied to plasma physics by \citet{BrackbillBarnes1980}
proved to be not usable in the present context. This is due to the
fact that there result some spurious oscillations at the smallest
spatial scales, which strongly distort the spectrum \citep[see
  also][]{Kissmann2006}. This leads to an obvious deviation of the
spectrum in the dissipation range, which cannot be tolerated.

Therefore, we are using a constrained transport evolution of the
scheme in order to assure the solenoidality of the field. This method
proved to be highly efficient and does not produce numerical
artifacts. See also in \citet{EvansHawley1988} for a derivation of the
scheme. The resulting scheme was extensively tested by
\citet{Kissmann2006} and was proved to fulfil all the requirements
stated above.

\section{Implementation of the Driving Process}
\label{SecDrive}
As was discussed above the driving process used in this work has to
fulfil several important constraints: First the fluctuations have to
be at large spatial scales. Then, the fluctuations have to be
completely random, while the energy distribution over the different
scales has to satisfy some form of scale dependence. Apart from that
we only used a fully solenoidal velocity field to drive the
fluctuations. Finally the overall energy of the fluctuations to be
added to the velocity field has to correspond to some physical values
(the supernova energy input rate in our case).

In this section we discuss the specific procedure that results in
the random fluctuation field. We start by giving the random
fluctuations in wave-number space -- thus, we can restrict the
fluctuations to large spatial scales. For this we worked closely along
what is suggested in literature \citep[see
especially][]{vestuto_etal_2003, christensson_etal_2001,
StoneEtAl1998}. The idea is to introduce fluctuations of random
amplitude, with the amplitudes distributed according to a normal
distribution on each of the input scales. The dependence of the
fluctuation energy on the spatial scale is at the same time fixed by
including a scale dependence of the variance $\sigma_{\delta
\vec{u}}^k$ of the normal distribution. For a normal distribution
around zero this just determines the width of the Gaussian $G_{\delta
\vec{u},k}$. For the velocity fluctuations $\delta \vel$ this means we
have:
\begin{equation}
  \left(\sigma_{\delta \vec{u}}^k \right)^2
  = \int \delta \vec{u}^2 G_{\delta \vec{u},k}
  d^3 \delta u
  \equiv \left< \delta \vec{u}(\vec{k})^2 \right>
\end{equation}
That is, in this case the variance is equivalent to the second-order
moment of the velocity fluctuations at scale $k$. Thus, we fix the
average form of the initial spectrum by prescribing the scale
dependence of $\sigma_{\delta \vec{u}}^k$.

The next step is to transform the resulting field to configuration
space. There we assure its solenoidality by taking the curl of the
random field. This operation, however, also changes the slope in
wave-number space. In particular we have:
\begin{equation}
  \label{EqCurlTransf}
  \nabla \times \vec{A}(\vec{x})
  \quad \longrightarrow \quad
  \vec{k} \times \vec{A}(\vec{k})
\end{equation}
This means that for the omni-directional spectrum there results an
additional factor of $k$ for the spectral slope of the velocity power
spectrum.  An omni-directional energy spectrum is eventually obtained
from the angle-integrated form of the square of the velocity -- that
is the velocity spectra resulting from the above differential
operation have to be squared and multiplied by an additional factor of
$k^2$ (the latter resulting from the angle integration). With all this
in mind it is possible to set the wave-number dependence of the
variance in a way as to provide any desired spectral slope. Fixing the
wave-number dependence of the standard deviation, thus, yields a
spectral slope of the form:
\begin{equation}
  E(k)_{\mathrm{Init}} = k^2(k\ \sigma_{\delta \vec{u}}^k)^2
\end{equation}
From the corresponding inverse relation we find that if a random
field with an average power law index of $-s$ is desired, the variance
has to fulfil the relation:
\begin{equation}
  \sigma_{\delta \vec{u}} \propto k^{-s/2-2}
\end{equation}
Thus, the full procedure is to give the desired driving in wave-number
space, at the same time restricting the spectrum only to small
wave-numbers. After this the fluctuations are transformed into
configuration space, where we only use their solenoidal part. Finally
we determine the energy and the momentum added to the numerical domain
from the resulting field. The former of these is the reduced to zero,
whereas the latter is normalised to the desired overall input energy.

\bibliography{KissmannEtAl2008MNRAS}

\label{lastpage}
\end{document}